\documentstyle[twoside,fleqn,espcrc2,epsfig]{article}

\newcommand{\be}{\begin{eqnarray}}
\newcommand{\ee}{\end{eqnarray}}

\newcommand{\AmS}{{\protect\the\textfont2
  A\kern-.1667em\lower.5ex\hbox{M}\kern-.125emS}}

\title{Unitarization of Total Cross Section and Coherent Effect in
pQCD}

\author{C.S. Lam\address{Department of Physics, McGill University\\
3600 University St., Montreal QC, Canada H3A 2T8\\Email:
Lam@physics.mcgill.ca}}

\begin{document}

\begin{abstract}
A formula to unitarize the leading-log BFKL-Pomeron amplitude
is derived using a coherent property of two-body collision in the peripheral region.
This procedure also allows
an algebraic characterization of the Reggeon in QCD based on color, instead of the
total angular momentum of the gluons being exchanged.
\end{abstract}

\maketitle

\section{BASIC IDEA}
I want to discuss a method to unitarize the leading-log BFKL Pomeron \cite{BFKL},
so that the growth of total cross section $\sigma_T(s)$ with energy $\sqrt{s}$
obeys the Froissart bound $\ln^2s$. The idea is very simple. Total cross
section grows because the Yukawa cloud of the colliding particles overlap even
at large impact parameters.
At low energy the clouds are so tenuous and transparent that
they do not contribute to the cross section. But as energy increases, the rarified overlap
 may contain sufficient energy to produce there a gluon jet, or a pair. When
that happens the cloud becomes opague and the effective radii inrease. This continues
until shadowing correction becomes important, at which time the rise is dampened and 
the Froissart bound is reached. This idea is familiar with potential scattering,
where Born approximation, if large, must be supplemented by higher-order corrections to restore
unitarity. In the present context, instead of 
potentials we must talk about interacting Reggeons, but the idea is the same. 
 The technical question is how to do so in QCD, where the basic entities are quarks
and gluons carrying non-commuting colours, rather than Reggeons. Clearly a characterization
of multiple Reggeons in terms of quarks and gluons are needed. We shall show that this 
can be derived with the help of 
an $s$-channel
factorization property, which also leads to an eikonal form 
for the amplitude which includes shadowing correction needed for the Froissart bound to be
obeyed.
Moreover, such a factorization is almost synonomous to the statement that the
state in the peripheral region in the collision is coherent. 

\section{COHERENCE}

 Consider $n$ bosons being emitted
from an energetic source with vertex factors $V_i$, as in Fig.~1(a). 
After Bose-Einstein symmetrization, {\it i.e.,}
summation over the $n!$ permuted diagrams, it can be shown \cite{LAM} that
each diagram can be factorized into a product of {\it quasi-particle} amplitudes.
A quasi-particle is made up of any number $p$ of gluons, which couples to the source
by the nested commutator $[V_1,[V_2,[V_3,\cdots,[V_{p-1},V_p]\cdots]]]$.
Therefore it is a colour-octet object
just like a single gluon. Indicating factorization by a vertical bar, 
and a permuted diagram by the gluon lines in the order they appear,
here are three examples showing how some $n=8$ diagrams factorize: $[1|2|3|4|5|6|7|8],
[5731|2|84|6], [1|2|3|854|76]$. The general rule is that a vertical bar is put
after a number iff no number to its right is smallest than it. Let
$C_m^\dagger$ be the operator creating a quasi-particle with $m$ bosons from the
vacuum state $|0\rangle$, then the bosons of these three examples are in the states
$(C_1^\dagger)^8|0\rangle, C_4^\dagger C_1^\dagger C_2^\dagger C_1^\dagger |0\rangle$,
$(C_1^\dagger)^3C_3^\dagger C_2^\dagger |0\rangle$, respectively. Summing over $m$,
the bosons emitted by the energetic particle is seen to be in a coherent state 
$\exp(C_1^\dagger+C_2^\dagger+C_3^\dagger+C_4^\dagger+\cdots)|0\rangle$.
Thus factorization and coherence are two sides of the same coin.
It is however important
to emphasize that this factorization occurs in the $s$-channel, so it is very different
from the usual factorization between short and long distances which occurs in the $t$-channel.
Moreover, the collection of all single quasi-particles turns out to be nothing but
a Reggeon, so the natural appearance of quasi-particle through factorization above can
be regarded as an algebraic characterization of the Reggeon.

\begin{figure}[htb]
\vspace{-5pt}
\framebox[75mm]{\rule[-121mm]{0mm}{43mm}}
\includegraphics{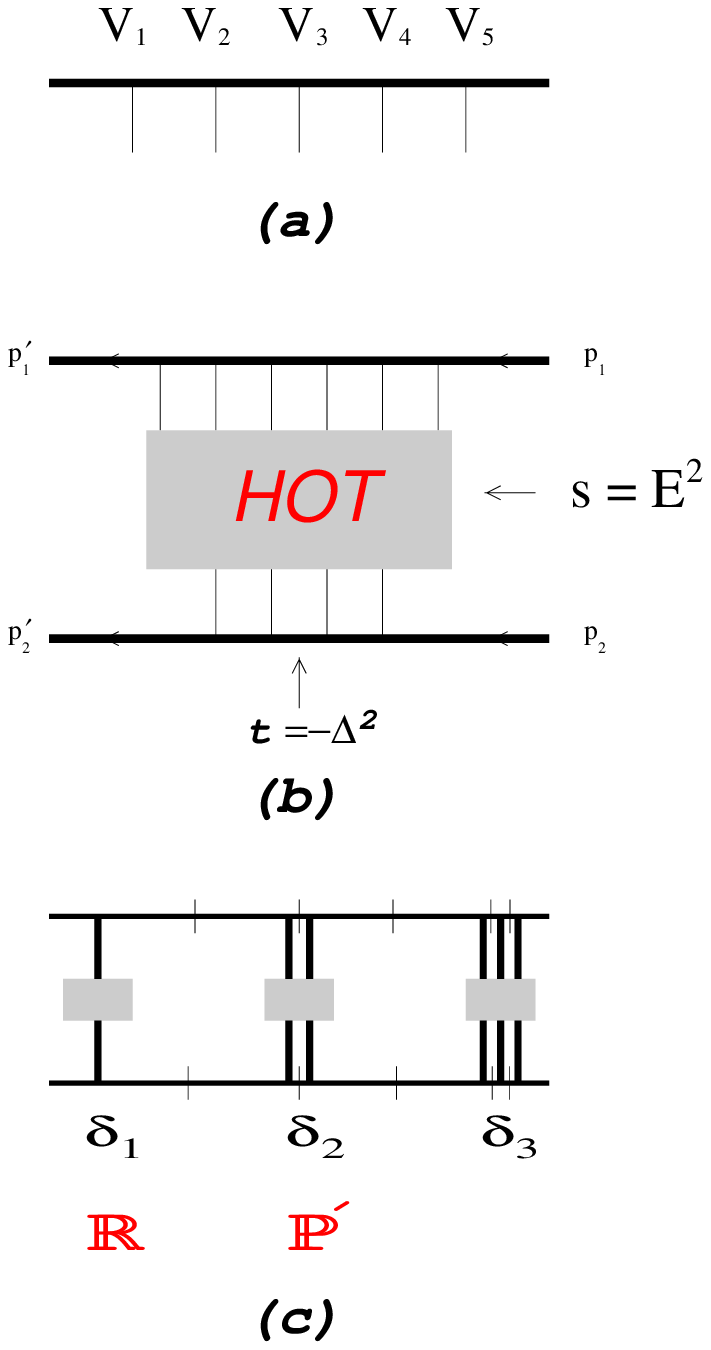}
\caption{(a) A tree diagram with vertices $V_i$; (b) a two-body
scattering diagram; (c) factorization of the two-body amplitude. Thick
vertical lines represent quasi-particle exchanges.}
\label{fig:l}
\end{figure}

\section{UNITARIZATION}
Consider a two-body scattering diagram, Fig.~1(b). The central region is hot and highly incoherent. However, as noted in the last section,
the peripheral tree amplitudes associated with the two energetic particles
are factorizable and the peripheral regions are `cool' and coherent. The precise 
manner factorization takes place depends on the numbering of the gluon
lines, which can be specified at will once and for all
for every set of permuted diagrams. We can use
this property to factorize the scattering amplitude as follows \cite{DKL}. 
Suppose the central region
of Fig.~1(b) falls into $k$ disconnected parts once the two energetic particles are removed.
For example, $k=3$ for Fig.~1(c). Then by suitably choosing the numbering of gluon lines,
we can always produce vertical bars (cuts) between every
disconnected components, as shown. Whether cuts occur inside each disconnected component depends
on the particular permutation of the lines within that component. Fig.~1(c) illustrates 
the case when no (one, two) cut occurs
in the first (second, third) disconnected component. The gluons
between cuts form a single colour-octet quasi-particle, indicated in Fig.~1(c) 
by a single thick line.
Thus the first (second, third) disconnected component is made up of the 
exchange of one (two mutually interacting, three mutually interacting)
quasi-particle(s). One can show in the
{\it extended leading-log approximation}, which keeps only terms with the lowest
power of the fine structure constant $\alpha_s$ for fixed $\xi\equiv\alpha_s\ln s$
and for amplitudes with a fixed number of quasi-particles exchanged,
that (A). The number of quasi-particles emerging from the bottom line of each disconnected
component is equal to the number of top; those that are not equal are subleading in the
approximation; (B). For fixed $\xi$ the amplitude with $m$ quasi-particle exchanged is
of the form $\alpha_s^md_m(\xi)$. 
Thus for example the amplitude in Fig.~1(c) is of order $\alpha_s^6$; and (C). The
non-commuting parts of the colour matrices between different irreducible parts contribute
to subleading terms so that we may effectively assume them to commute within the extended
leading-log approximation.

Summing up all possible diagrams we get the total amplitude in the extended leading-log
approximation to be
\be
{\cal A}(s,b)&=&1-\exp(2i\delta(s,b)),\\
\delta(s,b)&=&\sum_m\delta_m(s,b),\ee
where $\delta_m(s,b)$ is the contribution of a disconnected part with $m$ mutually
interacting quasi-particles being exchanged; each quasi-particle 
is a colour octet made up of
any number of gluons with arbitrary complexity.  We see from this eikonal form that $\delta(s,b)$ is simply
the phase shift at energy $\sqrt{s}$ and impact parameter $b$.

When the phase shift is small, we may replace the amplitude ${\cal A}(s,b)$
by its Born approximation ${\cal A}'(s,b)$. Using a subscript to denote the
number of quasi-particles being exchanged, we have ${\cal A}'_1=-2i\delta_1$
and ${\cal A}'_2=-2i\delta_2+2\delta_1^2$. The former is of order $\alpha_s$
and the exchanged object is a colour octet. This is nothing but the familiar
Reggeon amplitude in the leading-log approximation. Thus quasi-particles
turn out to be Reggeons in this context, but note that the concept of a
quasi-particle is far more general than that of a Reggeon.
The latter Born amplitude is of order $\alpha_s^2$
and two colour-octets are being exchanged. This contains (at least) a
colour singlet and a colour octet. The octet amplitude is negligible compared to
${\cal A}'_1$ and will be neglected. The singlet amplitude is nothing but
the leading-log BFKL Pomeron amplitude. 

The amplitude $A(s,\Delta)$ with momentum transfer $\Delta$ 
is given by the Fourier transform, and the total
cross section $\sigma_T(s)$ is given by the optical theorem, to be
\be
A(s,\Delta)&=&2is\int d^2b\ e^{i\Delta\cdot b}{\cal A}(s,b),\\
\sigma_T(s)&=&s^{-1}{\rm Im}A(s,0).\ee

\section{PHENOMENOLOGY}

These formulas can also be used to compute the energy variation of total
cross sections. Assuming the functions $d_m(\xi)$ for $m>2$ not to be substantially
larger than their counter part at $m=1$ or $m=2$,
the contribution from $\delta_m$ for $m>2$ can be ignored and $\sigma_T(s)$ can be computed
from $\delta_1$ and $\delta_2$, which in turn can be obtained from
the leading-log Reggeon amplitude ${\cal A}'_1$ and the BFKL amplitude ${\cal A}'_2$.

The former is completely known but unfortunately the latter is only partially known.
However, from direct perturbative calculations, scattering amplitudes are known to
8th orders \cite{CHENG}. We can extract from such calculations phase shifts
to the same order, and use it to calculate the total cross section in QCD.
Unitarity is guaranteed, but since we have not used the full Reggeon and BFKL
phase shifts to all orders, the calculation may not be numerically accurate.
We refer the readers to Ref.~\cite{DKL} for the result of the calculation and
comparision with the experimental data \cite{ZEUS}.

\section{CONCLUSION} 

A unitary formula
for total cross section is derived by making use of the 
coherent property in the peripheral region of the collision. The quasi-Particle 
so emerged turns out to be nothing but Reggeon fragments. In this way, not only
Froissart bound is guaranteed to hold, but an algebraic characterization of the
Reggeon is obtained.

\end{document}